\documentclass[USenglish,twocolumn]{article}

\usepackage[utf8]{inputenc}
\usepackage[big]{dgruyter}
\usepackage{natbib}
\usepackage{graphicx}

\begin{document}

  \articletype{Research Article{\hfill}Open Access}

  \author*[1]{Elena Shablovinskaya}

\author[2]{Viktor Afanasiev}

  \affil[1]{Special Astrophysical Observatory of the Russian Academy of Science, Nizhnii Arkhyz, 369167, Russia, E-mail: e.shablie@yandex.com}

  \affil[2]{Special Astrophysical Observatory of the Russian Academy of Science, Nizhnii Arkhyz, 369167, Russia, E-mail: vafan@sao.ru}

  \title{\huge Active Galactic Nuclei in polarized light}

  \runningtitle{AGN in polarized light}


  \begin{abstract}
{Due to the compactness active galactic nuclei (AGNs) are still unresolved with optical observations. However, structure and physical conditions of the matter in their central parts are especially important to study the processes of the matter accretion to supermassive black holes and eventually these investigations are essential to understand the galaxy evolution. Polarization contains information about the interaction of electromagnetic waves with the environment and provides information about the physical processes in the central regions of the AGNs that could not be found with the help of other optical observations. In this paper the importance of applying polarimetry methods to the study of geometry, kinematics, and physical processes in active galactic nuclei (AGN) in polarized light is discussed. An overview of the mechanisms of polarization formation, their connection with different structures and scales are provided. Also, we overview the polarimetric investigations based on different assumptions that are done using the observations conducted in Special Astrophysical Observatory of Russian Academy of Sciences.}

\end{abstract}

  \journalname{Open Astron.}
\DOI{https://doi.org/10.1515/astro-2019-0020}
  \startpage{213}
  \received{Jul 28, 2019}
  \accepted{Nov 27, 2019}

  \journalyear{2019}
  \journalvolume{28}

\maketitle
\section{Introduction}
Active galactic nuclei (AGN) are bright, compact regions emitting up to 90\% of the entire galaxy energy. Now the high rate of energy release in AGN is associated with the accretion processes onto the supermassive black hole (SMBH). The activity of nuclei is manifested by a number of extraordinary observed properties. Even in the first observations of AGN, the variability in all spectral ranges was observed both on long-term scales of few years (the first works in optics were Sharov\& Efremov 1963, Matthews \& Sandage 1963, Smith \& Hoffleit 1963a,b; in the radio band Dent 1965) and few hours (Wagner \& Witzel 1995). Atypical for normal galaxies emission lines in AGN spectra are also variable (see e.g. Ili{\'c} et al. 2015, Shapovalova et al. 2016, Shapovalova et al. 2017, Shapovalova et al. 2019). Moreover, the amount of observational data on changing-look quasars where broad lines have appeared or disappeared in the spectrum has recently increased (e.g., Khachikian \& Weedman 1971, Cohen et al. 1986, Goodrich 1989 and more recent e.g. MacLeod et al. 2019 and even in polarized light e.g. Hutsemékers et al. 2019). In general, active galaxies are brighter than normal ones, which is especially evident in significant energy excesses in radio, IR, UV bands and at higher frequencies. The spectral energy distributions (SED) of AGN are now considered to be composed of many components (see e.g. Collins et al. 2014), and the major part of radiation is contained in the optical-gamma bands produced in the central region of AGN with an average size of only 1 pc. 

However, an important property, which will be in a central place in this article, will be the polarization of active nuclei. As soon as non-thermal synchrotron spectra were detected in AGN, it was assumed by analogy with the radiation of the Crab Nebula the radiation of AGN should be polarized (Dibai \& Shakhovsky 1966). Soon linear polarization was indeed discovered (e.g. Efimov et al. 1979, Hough et al. 1987). As it was predicted by Antonucci (2002) and is absolutely clear today polarization provides additional parameters of the radiation of the studied source and is essential in AGN investigation. The central regions of AGN are now spatially unresolved in optical band, and it is impossible to study them using direct images, so the models of the AGN inner structure have only indirect evidences. Yet, using interferometric methods in radio and IR bands as GRAVITY or EHT it was recently obtained that the principle assumptions of AGN theory are correct such as the accreting SMBH or rotating matter in broad line region (BLR) (Castelvecchi 2019 and Gravity Collaboration et al. 2018, respectively). At the same time, polarization preserves information about both the spatial structure of the nucleus and the physical properties within.

This paper is a short review of polarimetric studies of AGN in the optical range, carried out by observations on the 6-m BTA telescope at the Special Astrophysical Observatory of the Russian Academy of Sciences (SAO RAS), and aims to emphasize the importance of AGN studies in polarized light. 

The paper is organized as following. In Sec. 2 we describe the mechanism producing polarization in AGNs in general. Then the individual cases (continuum polarization, broad line polarization and its rapid variability) are analyzed in Sec. 3, 4 and 5, respectively. Finally, in Sec. 6
we outline our conclusions.

\section{Polarization in AGNs}

\begin{figure*}
    \centering
    \includegraphics[scale=0.6]{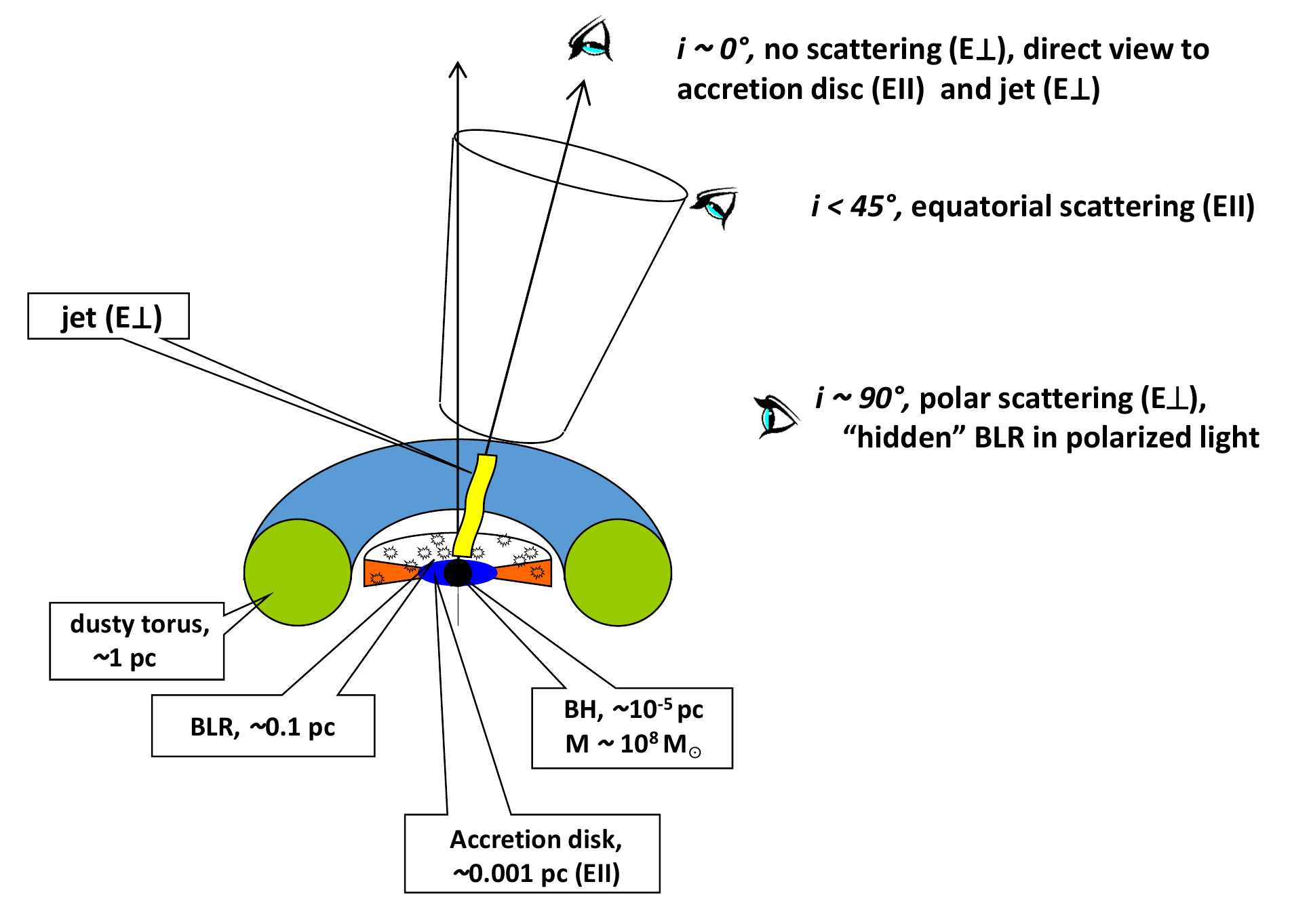}
    \caption{"Unified Model in polarized light". It is clear from the figure that the AGN orientation relative to the observer has a strong influence on the dominant polarization mechanism and its orientation. All major AGN components and their sizes are denoted. }
    \label{PUM}
\end{figure*}

As it is known, radiation becomes polarized if it passes through an anisotropic medium. Anisotropy can be caused by various mechanisms, which are usually divided into reflection and transmission processes. The degree of polarization of the reflected radiation is high (up to 100\%), while it usually causes a smaller polarization when passing through the medium. Concerning active nuclei, it is more convenient to divide the polarization mechanisms into internal (inside the central parsec) and external. 

The main internal mechanisms of polarization are:

\begin{itemize}
    \item[(i)] the polarization of radiation due to the influence of the magnetic field near the SMBH event horizon within the frames of general relativity (the Lense-Thirring effect);
    \item[(ii)] the polarization of accretion disk radiation due to radiation transfer (Beloborodov 1998) and Faraday rotation in magnetized medium (e.g. Gnedin et al. 2006, Silant'ev et al. 2009);
    \item[(iii)] the optical jet synchrotron radiation;
    \item[(iv)] the scattering by electrons in a hot corona.
\end{itemize}

The internal mechanisms of polarization are summed up, and it is impossible to separate them unambiguously though they have the different direction and they vary on different time scales.

Among the external mechanisms are only two of them:

\begin{itemize}
    \item[(i)] the equatorial scattering at the dusty torus;
    \item[(ii)] the polar scattering at the ionization cone.
\end{itemize}

These mechanisms are clearly observed in Seyfert galaxies (Sy), which are classified into two types due to the features in the spectrum: in Sy 1 there are both broad and narrow emission components, whereas in Sy 2 there are only narrow ones. Antonucci (1982) found that two types of Sy galaxies are also characterized by different orientation polarization: in Sy 1 polarization tends to be parallel to the radio axis, and in Sy 2 to be perpendicular. Within the Unified model (UM, Antonucci 1993; Urry \& Padovani 1995) these observational properties are explained by the orientation of the active nuclei relative to the observer, so the dusty torus can obscure the central regions or not, and therefore this determines the dominant scattering mechanism. Therefore, the different polarization mechanisms dominate in different AGN types. The "unified model in polarized light" is presented in Fig. \ref{PUM}. It describes the types of polarization mechanism depending on the AGN orientation relative to the observer. The AGN orientation relative to the observer has a strong influence on the dominant polarization mechanism and its orientation. At different angles an observer could see different AGN components. At the angles close to $45^\circ$ - i.e. in case of Type 1 AGNs - the central parsec are observed (accretion disk, BLR, region of the jet formation) together with the dusty torus. So, in their spectra both broad and narrow lines are seen. Note here that within the Type 1 AGNs one can also find significant difference as it is concluded in the paper by Popović (2018). At lower angles closer to the equatorial plane - i.e. Type 2 AGNs - the dusty torus obscure the inner regions from the observer and only narrow lines are detectable as NLR is located much farther. The detection of broad emission lines in polarized light in the galaxy NGC 1068, previously considered a typical representative of galaxies of Sy 2 type (Miller et al. 1991), was a good proof of the described assumption of obscuration and the UM in general.  Then, observing the AGN at very high angles one watch approximately into the jet cone, where the radiation is so energetic that its contribution "swamp"\ the details. 

Moreover, the polarimetric studies of NGC 1068 were conducted by {\it HST} with the Faint Object Camera in the ultraviolet and with the WF/PC-I in the visual (Capetti et al. 1997 and references therein). Observations have shown the polarization changing across the field. A few years after Kishimoto (1999) reanalyzed the {\it HST} polarimetric data and constructed a model under the assumption of light coming from a point source and scattering at clouds (in this case, NLR), and the two-dimensional polarization distribution was de-projected into a three-dimensional matter distribution in the AGN.

According to the above considerations it can be seen that being associated with the physical state of matter and its spatial distribution polarization becomes an essential tool to study the spatially unresolved areas of AGN. To illustrate this, the results of AGN studies in polarized light in the continuum and broad emission lines, as well as studies of the rapid polarization variability, will be presented below.

Note here that the Stokes parameters of polarization are used in paper. This formalism was introduced by Stokes (1852). Within the frames of this formalism $I$ parameter is the total object intensity, $Q$ and $U$ describes the linear polarization and $V$ is for the circular one. to find a mathematical description see (Walker 1954).

\section{Polarization in continuum}

It is known that in AGN polarization depends on the wavelength, and it s due to different mechanisms of polarization formation. Webb et al. (1993) did the first attempt to separate the different mechanisms of continuum polarization, one of which could be an accretion disk. However, even with the anisotropic distribution of the hot gas in the accretion disk and hence the anisotropic Thomson scattering, the observed polarization should not depend on the wavelength. The situation changes if there is a magnetic field in the accretion disk. Gnedin \& Silant'ev (1984, 1997) found that for weak ($B < 10^3$ G)\footnote{The expected theoretical range for magnetic field strength in SMBH accretion disks is $10^{-2}-10^4$ G (see e.g. Silant'ev et al. 2009 and references therein).} magnetic fields in the accretion disk, optical anisotropy occurs due to the Faraday rotation of the polarization plane at the photon free path, which leads to a strong dependence of the degree of linear polarization on the wavelength $\lambda$. Spectropolarimetric observations of a sample of AGN at the 6-m BTA telescope (Afanasiev et al. 2011) shown that the dependence $P(\lambda) \propto \lambda^n$ exists, where $P$ is the polarization degree and $\lambda$ is wavelegth, and the polarization is formed at small spatial scales – as it will be shown later, less than the BLR size, and can be associated with the processes in the accretion disk. Silant'ev et al. (2009) shown that if the temperature changes along the radius according to the $T(R) \propto R^{-p}$ law (as for the standard Shakura-Sunyaev disk (Shakura \& Sunyaev 1973)) and the magnetic field strength also changes radially $B(R) \propto B_H (R_H/R)^s$, where $R_H$ and $B_H$ are the event horizon radius and the magnetic field strength at it, respectively, then the degree of polarization will change with the wavelength as $P \propto \lambda^{(s/p-2)}$. Comparing this ratio with the observed one, the magnetic field strength ranging from 0.37 to 300 G was obtained for 14 AGN. As it was discussed in the paper (Afanasiev et al. 2011) the magnetic field strength decreases with
increasing increasing SMBH mass, and it is in a agreement with the magnetic coupling model.

Besides, it is possible to determine the radiative efficiency of the accretion disk, depending on the spin of the SMBH. Du et al. (2014) shown that the radiative efficiency could be calculated knowing mass, luminosty, Eddington ratio and the inlination angle of AGN. The determination of inclination angle is the hardest problem here as the systems are not optically resolved. Afanasiev et al. (2018) shown that obtaining the polarization of the source and comparing it with the predicted by Sobolev–Chandrasekhar theory the inclination could be found. Such calculations were made for 47 AGN, and 90\% of the sample show spins >0.9, which corresponds to Kerr black holes. 

Also, the polarization in the continuum varies with the amplitude of about 1-2\%. The spectropolarimetric monitoring was conducted at BTA telescope for Sy 1.5 Mrk 6 (Afanasiev et al. 2014) and Sy 1 3C 390.3 (Afanasiev et al. 2015). According to the obtained observational data it was found for Mrk 6 that the delay in broad H$_{\alpha}$ line relative to the continuum at 5100\AA\ is 22 days, which corresponds to the size of the BLR region - 22 lt days (it is consistent with the reverberation mapping results by Bentz \& Katz (2015)). As well, the polarized continuum at 5100\AA\ shows only 2 days delay\footnote{Yet the observations sampling was greater than 2 days (there were 13 epochs at $\sim$1000 days, the method of interpolated cross-correlation function was applied. The linear interpolation was used. Comparison of the linear interpolation with interpolation by exponential continuum model shown that the results were corrected.}, that corresponds to the size by an order smaller than BLR. A similar result was obtained for 3C 390.3: the size of the BLR in H$_{\alpha}$ was 120 lt days (for H$_{\beta}$ was 2 times less, 60 lt days), and the size of the region where polarized continuum forms – 10 lt days. Thus, the region of the polarized continuum, approximately 10 times smaller than BLR, and it is consistent with the idea of the polarized radiation coming from the accretion disk.

\section{Polarization in broad lines}

\begin{figure}
    \centering
    \includegraphics[scale=0.8]{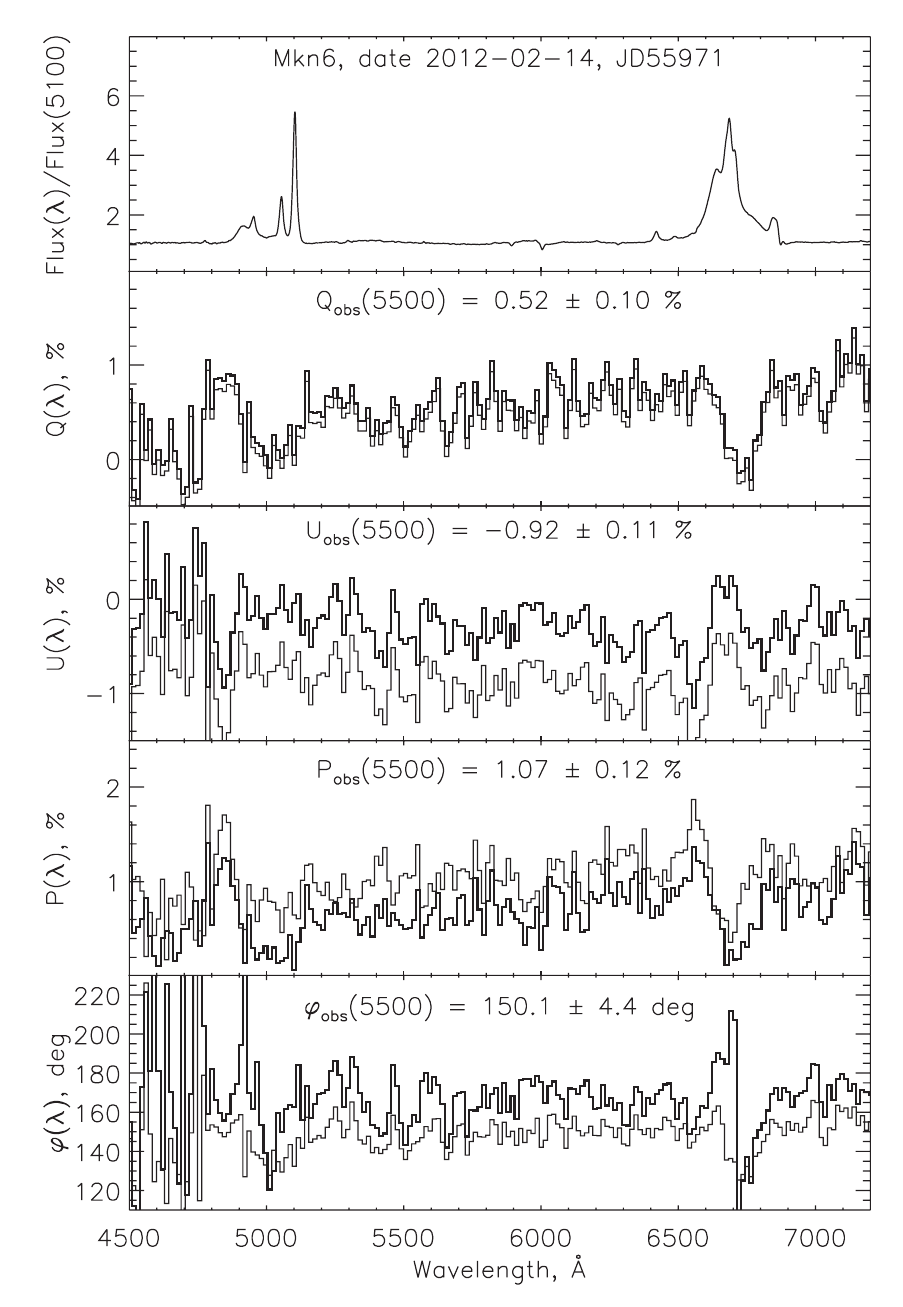}
    \caption{The spectrum and its polarization parameters observed on 2012 Feb 14. From top to bottom: the observed spectrum of Mrk 6, parameters Q and U and corresponding degree $P$ and angle $\varphi$ of polarization. The thin lines represent observed and the thick lines represent corrected spectra for the ISM polarization. The figure is taken from (Afanasiev et al. 2014).}
    \label{mrk6}
\end{figure}

Discussing in this section, Sy 1 galaxies are characterized by the broad emission lines corresponding to the allowed and semi-allowed transitions (as the Balmer series, MgII, CIV, CIII], etc.) in their spectra. It is considered that the lines are formed in the BLR region located at an average distance of 0.01-0.1 pc from the central source. The broadening of the lines is due to the Keplerian rotation of the BLR clouds as a thick disk and shows the velocities up to $10^4$ km/s. Due to the small spatial sizes BLR could not be optically resolved even in close AGN, so only indirect methods are used to study it. The most popular is the reverberation mapping (Blandford \& McKee 1982), which allows us to determine the distance between the BLR and the central source (more precisely – the accretion disk, where exciting radiation is formed) by the delay between the continuum radiation and re-emitted one in the emission line. The BLR stratification was revealed (Peterson 1994; Baldwin et al. 2003) and the characteristic sizes in more than a hundred galaxies were determined. The leading role now plays SDSS-RM project (Shen et al. 2015) aimed to measure time delays for 849 galaxies at the redshifts up to 4.5. Here we will note that the most important achievement of RM programs is recently the relation of AGN luminosity versus the BLR size (Bentz et al. 2009) that indirectly proves the model of the matter accretion onto SMBH. To find more details on RM see (Bentz 2015) However, the reverberation mapping does not allow to conclude about the character of the clouds motion, and also depends on the angle of the system inclination, not determined directly. 

Spectropolarimetry is a powerful tool for studying the kinematic properties of BLR. In Sy 1 AGNs, the dominant polarization mechanism in the lines is external equatorial scattering at the dusty torus (the internal mechanisms make a small contribution, and the emission line can be assumed unpolarized before the scattering). Smith et al. (2005) found that due to the Keplerian rotation of the clouds in BLR within equatorial scattering the polarization angle $\varphi$ has an S-shaped profile depending on the wavelength $\lambda$ along the emission line. The example of polarized observations of H$\alpha$ line in Mrk 6 is demonstrated in Fig. \ref{mrk6}. A more detailed simulation using the {\it STOKES} package (Marin 2018) showed that the equatorial scattering does not affect the line profile in total flux, but in polarized light the two-hump structure and the dramatic change of the polarization angle along the profile become visible. Afanasiev et al. (2014) shown that if the motions are Keplerian ones, then the velocity can be expressed as:
$$
\log(v_{i}/c) = a - b\cdot \log(\tan(\varphi_i)), 
$$
$$
a=0.5 \log \Big(\frac{GM_{BH} \cos^2(\theta)}{c^2R_{sc}}\Big),
$$
where $v_i$ and $\varphi_i$ is the velocity and the polarization angle for the definite $\lambda_i$,$c$ is the speed of light, $G$ is the gravitational constant, $M_{BH}$ is a central black hole mass, $\theta$  the angle between the disc and polarization plane (see Fig. 1 in (Afanasiev \& Popovi\'c 2015)) and $R_{sc}$ is the distance between the central source and the scattering region. Thus, observing clouds rotating at Keplerian orbits, the obtained dependence $\log(v_{i}/c)$ vs. $\log(\tan(\varphi_i))$ must be linear, which was discovered firstly by modeling, then by observations in the broad H$_{\alpha}$ line in Mrk 6 (Afanasiev et al. 2014) and in 3C 390.3 (Afanasiev et al. 2015), and for for 35 Sy 1 galaxies (Afanasiev et al. 2019). 

It is important to note that the parameter $a$ obtained by linear approximation of the dependence $\log(v_{i}/c)$ vs. $\log(\tan(\varphi_i))$ also depends on two important parameters of AGNs – the radius of the scattering region, or dusty torus, $R_{sc}$ and the mass of the central SMBH $M_{BH}$. This fact is the basis of the method of the SMBH mass determination in Sy 1 galaxies with spectropolarimetric observations, described by Afanasiev \& Popovi{\'c} (2015). 

Let’s compare the proposed method of the mass determination with reverberation mapping. Assuming the gas is virialized and the size of the BLR is identically equal to the observed signal delay $R_{BLR}=c\tau$, the relation $M_{SMBH} = f {v^2 R_{BLR}}/{G}$ could be used. This estimation is affected by a number of assumptions. The observed gas velocity depends on the inclination angle of the system: $v = v_{obs} / \sin(i)$, and the angle $i$ is not determined from direct observations, since BLR is spatially unresolved. The factor $f$ is also determined empirically. The coefficient $f$ is considered to describe the inclination angle of the system as well as the structure of BLR (e.g. Yu et al. 2019), however, there is no clarity on this issue yet. Thus, on the one hand, reverberation mapping has a significant advantage due to the fact that it uses the velocities at scales much closer than can be obtained from star velocity dispersions. However, a large number of unknown parameters are used for the estimation, which is why the accuracy of the mass measurement lies within the order of magnitude. In addition, to measure the key parameter – the signal delay between the continuum and the line – it requires a long series of observational data, which is implemented for a small number of galaxies and only at small redshifts.

At the same time, mass estimation by spectropolarimetry requires only 1 observation epoch, from which the dependence of the logarithm of the velocity on the logarithm of the polarization angle is obtained. As it can be seen from the expression for $a$, the only value expected additional measurements is the dusty torus size $R_{sc}$. Now there are two ways to determine it: IR reverberation mapping (Minezaki et al. 2004; Suganuma et al. 2006; Koshida et al. 2009, 2014; Mandal et al. 2018) and near-IR interferometry (e.g. Kishimoto et al. 2009, 2011). Comparison of $R_{sc}$ and $R_{BLR}$ (Afanasiev et al. 2019) showed that there is a linear dependence between these quantities, and $R_{sc}$ is on average 2 times more than $R_{BLR}$. However, the measurements of the dust sublimation radius is still a hard question as the theoretical predictions are not in agreement with IR observation. Also it should be mentioned here that except the difficulty with $R_{sc}$ estimation spectropolarimetric method of mass estimation could be applied only for the systems where the rotation motion dominates the radial velocities and out/inflows are weak. For more details of the method application see Savi\'c et al. 2018. 

It is important that the spectropolarimetric method of mass estimation does not depend on the orientation of the system. Then having two methods for determining the mass of BH, dependent and independent on the inclination, it is possible to determine the angle $i$. A sample of galaxies shows a linear relation between the inclination angle of the galaxy and the inclination angle of the BLR (Afanasiev et al. 2019). Accordingly, if we know the true angle of the BLR, the distribution of brightness on the disk could be estimated. From observations the distribution is close to the classical Sakura-Sunyaev accretion disk (Afanasiev et al. 2019).

\section{Short-term variability of polarization}

BL Lac type objects (or blazars)\footnote{Though the terms "BL Lacs" and "blazars" are not equal to each other, within this paper we would assume it interchangeably.} are a special type of AGNs with the jet directed at a small angle to the observer’s line of sight. Because of this orientation, the blazars look like star-like sources, their brightness can exceed the host galaxy luminosity by up to 4 mag. It is also known that their luminosity and polarization has a synchrotron origin and are highly variable even within the night. Impey et al. (2000) mentioned for the first time that the polarization of the blazar S5 0716+714 on scales of several hours walks on the $QU$-plane in arcs, that is equivalent to the rotation of the polarization vector. Shablovinskaya \& Afanasiev (2019) shown that on a longer time series the polarization vector in the blazar S5 0716+714 forms arcs and loops on the $QU$-plane, while the change of direction occurs within appoximately 1.5 hours. A similar, though less clear, picture is seen from the analysis of the 2014 microflash data described in Bhatta et al. (2015a,b) and for blazar BL Lac (Covino et al. 2015). Since observing BL Lac type objects one looks inside the jet and polarization appears due to the plasma rotation in the magnetic field, the rotations observed on the $QU$-plane correspond to the physical plasma motion in the jet. 
According to the model of the jet structure (Marscher et al. 2008), optical radiation is formed in a helical magnetic field at a distance of $<10^{-2}$ pc from the central source. Based on this fact, the observed plasma motion should be directed along the arcs, and the time of switching the direction of the polarization vector is equivalent to the size of the region where the radiation is formed. Since the region of the optical jet is not spatially unresolved, it is possible to use the polarization vector rotation on the $QU$-plane as a plasma motion test. Thus, similar arcs described by the polarization vector were found in the radio band polarization for the blazar CTA 102 (Li et al. 2018). However, as shown in (Shablovinskaya \& Afanasiev 2019), model of stable helical magnetic field cannot describe the observed rotation of the polarization, therefore the magnetic field precession was suggested, that for S5 0716+714 is approximately 15 days.

\section{Conclusion}

Polarimetry is a powerful tool to investigate the central regions of active nuclei that are unresolved by the direct observations. Being associated with the physical state of matter at different distances from the central SMBH polarization stores information about the inner structure of the nucleus. Observations of polarization carried out in recent years in SAO RAS allowed to obtain the following results.

\begin{itemize}
    \item     The polarization in {\it continuum} is produced in magnetized AD (0.001-0.01 pc) and depends on the magnetic field strength in accreation disk $B(R)$, SMBH mass $M_{SMBH}$ and spin. Spectropolarimetric investigations revealed that $B(R)=0.3-300$ G and spin $\sim$1 that corresponds to Kerr black holes.
    \item The polarization in {\it broad lines} resolves the gas kinematics in BLR (0.01-0.1 pc), and it leads to the more accurate SMBH mass estimation in Sy 1 galaxies, independent from the inclination angle. Also, comparison between these estimations and ones obtained by reverberation mapping the inclination angles were calculated for the number of galaxies and the luminosity distribution of the BLR disks with known inclination were shown.
    \item     {\it Short-term variability} of the polarization vector in BL Lac type objects marks the plasma kinematics inside the jet. Comparing the observational results with the models of the helical magnetic field in jet it was offered that the field is precessing and the region where the optical polarization is formed is of the 1.5 lt hour size. 
\end{itemize}

\footnotesize
\section*{Acknowledgments}
We sincerely thank the organizers of 12 Serbian Conference on Spectral Line Shapes in Astrophysics for the special session "Broad lines in AGNs: The physics of emission gas in the vicinity of super-massive black hole" organized in memory of Dr Alla Ivanovna Shapovalova who was our dear colleague and friend.

\bibliography{li}

\end{document}